\documentclass[letterpaper]{article} 
\usepackage{aaai25}  
\usepackage{times}  
\usepackage{helvet}  
\usepackage{courier}  
\usepackage[hyphens]{url}  
\usepackage{graphicx} 
\usepackage{natbib}  
\usepackage{caption} 
\usepackage{algorithm}
\usepackage{algorithmic}
\usepackage{comment}
\usepackage{amsfonts}
\usepackage{amsmath}
\usepackage{multirow} 
\usepackage{booktabs}
\usepackage[table,dvipsnames]{xcolor}
\usepackage{array} 

\usepackage{newfloat}
\usepackage{listings}
\DeclareCaptionStyle{ruled}{labelfont=normalfont,labelsep=colon,strut=off} 
\floatstyle{ruled}
\newfloat{listing}{tb}{lst}{}
\floatname{listing}{Listing}

\usepackage{tcolorbox}
\usepackage{pifont} 
\usepackage{float}      
\usepackage{caption}
\usepackage{algorithm}
\usepackage{algorithmic}
\usepackage{subcaption}

\usepackage{fontawesome5}
\usepackage{xcolor}

\definecolor{sbblue}{HTML}{4C72B0}
\definecolor{sborange}{HTML}{DD8452}
\definecolor{sbgreen}{HTML}{55A868}
\definecolor{sbred}{HTML}{C44E52}
\definecolor{sbpurple}{HTML}{8172B3}
\definecolor{sbbrown}{HTML}{937860}
\definecolor{sbpink}{HTML}{DA8BC3}
\definecolor{sbgray}{HTML}{8C8C8C}

\newcommand{\uparrowtrend}{{\color{sbblue}\faArrowUp}}
\newcommand{\downarrowtrend}{{\color{sborange}\faArrowDown}}
\newcommand{\stabletrend}{{\color{sbpurple}\faMinus}}
\newcommand{\correcttrend}{{\color{green}\faCheck}}
\newcommand{\wrongtrend}{{\color{red}\faTimes}}

\usepackage{makecell} 

\title{DEEP: A Discourse Evolution Engine for Predictions about Social Movements}
\author{
    Valerio La Gatta\textsuperscript{\rm 1}\equalcontrib, Marco Postiglione\textsuperscript{\rm 1}\equalcontrib, Jeremy Gilbert\textsuperscript{\rm 2}, Daniel W. Linna Jr.\textsuperscript{\rm 3}, Morgan Manella Greenfield\textsuperscript{\rm 4}, Aaron Shaw\textsuperscript{\rm 5}, V.S. Subrahmanian\textsuperscript{\rm 1}
}
\affiliations{
    \textsuperscript{\rm 1}Department of Computer Science and Buffett Institute for Global Affairs, Northwestern University\\
    \textsuperscript{\rm 2}Medill School of Journalism, Media, Integrated Marketing Communications, Northwestern University\\
    \textsuperscript{\rm 3}Northwestern Pritzker School of Law and Department of Computer Science, Northwestern University\\
    \textsuperscript{\rm 4}The Wall Street Journal\\
    \textsuperscript{\rm 5}Department of Communication Studies, Northwestern University\\
    vss@northwestern.edu
}

\usepackage{bibentry}

\begin{document}

\maketitle

\begin{abstract}
Numerous social movements (SMs) around the world help support the UN's Sustainable Development Goals (SDGs). Understanding how key events shape SMs is key to the achievement of the SDGs. We have developed SMART (Social Media Analysis \& Reasoning Tool) to track social movements related to the SDGs. SMART was designed by a multidisciplinary team of AI researchers, journalists, communications scholars and legal experts. This paper describes SMART's transformer-based multivariate time series Discourse Evolution Engine for Predictions about Social Movements (DEEP) to predict the volume of future articles/posts and the emotions expressed. DEEP outputs probabilistic forecasts with uncertainty estimates, providing critical support for editorial planning and strategic decision-making. We evaluate DEEP with a case study of the \emph{\#MeToo} movement by creating a novel longitudinal dataset (433K Reddit posts and 121K news articles) from September 2024 to June 2025 that will be publicly released for research purposes upon publication of this paper.

\end{abstract}

%

\section{Introduction}
Social movements (SMs) play an important role in supporting the UN's Sustainable Development Goals (SDGs)~\cite{salles2024social}. The \#MeToo movement alone has influenced the reporting of gender-related crimes~\cite{levy2023effects} and health effects linked to sexual harassment~\cite{o2018metoo}. SMs involved with the fossil fuel energy have had some success in the UK, Netherlands, and Poland~\cite{hielscher2022social}, and the rise of community choice movements in California ~\cite{smith2019energy}. In the Philippines, social movements supporting the environment have helped reduce pollution by companies~\cite{magno2017environmental} and protect indigenous peoples' rights~\cite{theriault2011micropolitics}. Even small progress supporting the UN's SDGs can have  a significant impact on millions of people, and SMs supportive of the SDGs attempt to do just that.

Our Social Media Analysis and Reasoning Tool (SMART)\footnote{\url{https://buffett.northwestern.edu/research/global-working-groups/ai-and-social-movements.html}{https://buffett.northwestern.edu/research/global-working-groups/SMART}} tracks SMs that are supportive of the UN's SDGs and makes information about them available to journalists. The 8 journalists when SMART/DEEP were started had four requirements. \textsf{(i)} They wanted to know about SMs that were rising in popularity in terms of volume (number of posts mentioning the SM). \textsf{(ii)} For such SMs, they wanted to understand the sentiments and emotions expressed in those posts and how they evolve over time. \textsf{(iiii)} They were interested in the relationship between key events (e.g., political events such as elections) and these SMs. \textsf{(iv)} As journalists want to publish articles on timely topics before other journalists do or the public broadly knows, they wanted to identify such SMs early. \emph{For these reasons, we designed our Discourse Evolution Engine for Predictions about Social Movements (DEEP) to forecast volume and emotional intensity so that journalists could look at forecasts about SMs of interest (e.g., a journalist in the Philippines might care more about SMs relating to fishing issues in her country than a journalist in Paris).} This way, a journalist can use the forecasts to hone in on SMs that are much talked about, and write about them \textit{early}, before others do. \emph{This paper is not about SMART, but it is discussed briefly for context.}

The main contributions of this paper are as follows. \textsf{(I)} We define the Discourse Evolution Prediction problem to predict the volume of posts about a specific SM \textit{S}, the sentiment about \textit{S}, and the emotional intensity of several emotions in posts about \textit{S}, $\Delta$ time units into the future. We then propose the DEEP framework to solve this problem. \textsf{(II)} We describe experiments that show how DEEP performs on data about the \#MeToo movement. DEEP achieves consistently higher precision, recall, and F1 on news than on Reddit, reflecting the clearer emotional signals in formal reporting. However, while short-horizon forecasts are most reliable for news, Reddit predictions improve with longer horizons, revealing platform-specific dynamics that enable accurate near-term detection in news and strong medium-term forecasting in social media discussions. \textsf{(III)} We provide a \#MeToo dataset consisting of  433,016 Reddit posts and 121,849 news articles from September 1, 2024 and June 28, 2025, along with an analysis involving key events from the Sean “Diddy" Combs case. The SMART/DEEP framework has been developed with consistent feedback from journalists, demonstrated to over 20 journalists to gain feedback, including at a small workshop in March 2025 to fine-tune the system\footnote{\url{https://buffett.northwestern.edu/news/2025/empowering-journalists-to-decode-social-movements-the-buffett-global-working-group-behind-new-ai-system.html}}. A list of the organizations involved is provided in the Appendix. We expect to make DEEP available to journalists in early 2026. As such, this paper falls within IAAI's Emerging Applications Track.

\section{DEEP Problem Formulation}
We formalize the DEEP problem as follows. The \emph{discourse state} $\mathcal{S}_t$ at time $t$ is a vector
$\mathcal{S}_t = (\mathbf{V}_t, \mathbf{E}_t, \mathbf{T}_t)$
where:
\textsf{(i)}
$\mathbf{V}_t \in \mathbb{R}^h$ represents the volume of posts/articles gathered at time $t$, \textsf{(ii)} $\mathbf{E}_t \in \mathbb{R}^d$ denotes emotional intensity across $d=28$ emotional dimensions averaged across all posts at time $t$, and \textsf{(iii)} $\mathbf{T}_t \in \mathbb{R}^m$ represents the distribution over $m$ topics of interest (e.g., politics, sports, entertainment). 

The \emph{Historical Trajectory} $\mathcal{H}_t$ is a sequence of discourse states up to time $t$: $
\mathcal{H}_t = (\mathcal{S}_1, \mathcal{S}_2, \ldots, \mathcal{S}_t)
$.

\emph{Key events} $\mathcal{K}_t$ are discrete occurrences that the journalist believes will significantly influence the dynamics of discourse about the phenomenon the journalist is interested in.  Formally, $\mathcal{K}_t = {k_1, k_2, \ldots, k_n}$ where $k_i = (\epsilon_i, t_i, \alpha_i)$. Each key event $k_i$ has a type $\epsilon_i \in \mathcal{E}$ (from a predefined event taxonomy), occurrence time $t_i$, and impact magnitude $\alpha_i \in \mathbb{R}^+$ (provided by a journalist).

\emph{The Discourse Evolution Engine Prediction (DEEP) Problem} is to estimate the future discourse state $\mathcal{S}_{t+\Delta}$ given the historical trajectory $\mathcal{H}_t$ and key events $\mathcal{K}_t$, i.e. to find a function $f$ such that:
\begin{equation}
\mathcal{S}_{t+\Delta} = f(\mathcal{H}_t, \mathcal{K}_{t:t+\Delta}, \boldsymbol{\theta}) \text{, }
\end{equation}
where $\mathcal{K}_{t:t+\Delta}$ represents key events the journalist expects to happen within the prediction window $\Delta$, and $\boldsymbol{\theta}$ denotes model parameters learned from historical data.

\section{Materials \& Method}

\begin{figure*}[t]
    \centering
    \includegraphics[width=\linewidth]{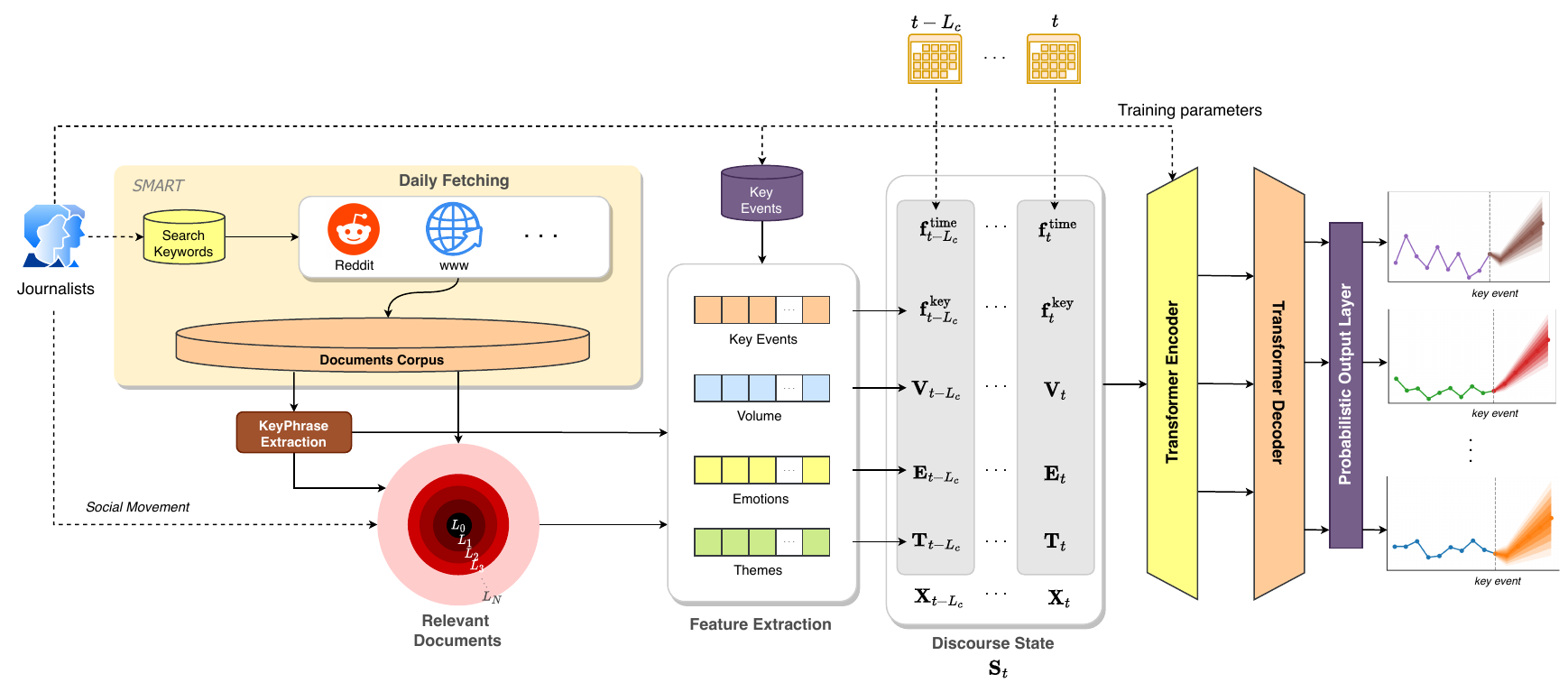}
    \caption{\textbf{Overview of DEEP.} Data is collected from Reddit and News using hashtag and semantic keyword extraction. Keyphrases determine document relevance to the input social movement. Feature extraction transforms text into structured representations across volume, emotions, themes, and key events. A TimeSeriesTransformer processes historical discourse through an encoder-decoder framework to generate probabilistic forecasts of future discourse states $S_{t+\Delta}$ with uncertainty quantification via Student-t distributions.}
    \label{fig:placeholder}
\end{figure*}

Our system uses a multi-stage pipeline that integrates data collection, feature engineering, and transformer-based forecasting. Figure \ref{fig:placeholder} presents an overview of our pipeline, illustrating the flow from raw data collection through feature extraction to predictive modeling. We now discuss these modules in depth\footnote{While the description focuses on the \#MeToo movement as the target SM, the methodology can be easily adaptable to any social movement.}. 

\subsection{Data Collection}

We identify data about a social movement through a corpus of content selected via keyword-based search and information retrieval processes that capture both explicit references to the movement and semantically related discussions. While our system is designed to support journalists predicting discourse evolution around specific social movements, this abstraction makes it sufficiently general to support other applications reducible to similar search and retrieval structures, including topic-based forecasting and thematic content analysis. For the \#MeToo movement, we extract data daily using the hashtag “\#MeToo'' alongside related search keywords identified by journalists (e.g., gender equality, women's voices, equal opportunities; the full list is reported in the Appendix).

The resulting comprehensive dataset consists of $433,016$ Reddit posts and $121,849$ news articles collected between September 1, 2024 and June 28, 2025. We now describe a multi-layer data extraction methodology we developed that captures both explicit mentions to the movement and related discourse.

\paragraph{Layer $L_0$} The first layer $L_0$ identifies documents that explicitly reference the target social movement $sm$ (i.e., $sm=$\#MeToo). Documents in this layer either were extracted via the “\#MeToo" keyword or contain it in their title or body text. Layer $L_0$ contains $3,103$ Reddit posts and $3,286$ news articles.

\paragraph{Layers $L_1$, $L_2$, $\dots, L_N$} To expand the dataset beyond documents containing direct mentions of $sm=$\#MeToo, we implemented a systematic approach to identify semantically related content. First, we extracted keywords and keyphrases for each document using KeyBERT ~\cite{sharma2019self} and Amazon Comprehend\footnote{\url{https://aws.amazon.com/comprehend/}}, respectively. We then selected all keywords that co-occur with $sm$ across our corpus and computed their frequency distribution. From this distribution, we identified keywords falling within the 99th percentile of co-occurrence, creating a keyword set that captures the core vocabulary associated with $sm$. We progressively relaxed thresholds to capture documents at varying degrees of semantic relatedness to the movement: layers $L_1$, $L_2$ and $L_3$ include documents containing at least 30\%, 20\% and 10\% of the high-salience keywords, respectively.

This tiered approach balances precision and recall, enabling the inclusion of documents that may not explicitly mention the target SM but contribute meaningfully to the broader discourse surrounding the movement. Table \ref{tab:dataset_summary} shows the number of Reddit posts and news articles retrieved within each layer.

\begin{table}[t]
\centering
\caption{Summary of Multi-layer Data Extraction Results}
\label{tab:dataset_summary}
\scalebox{0.8}{
\begin{tabular}{rccc}

\toprule
\makecell{\textbf{Layer}} & \textbf{Description} & \textbf{Reddit} & \textbf{News} \\ \midrule
$L_0$ & Direct \#MeToo mentions & $3,103$ & $3,286$ \\
$L_1$ & \makecell{Documents containing $\geq$30\% \\ of top 1\% co-occurring keywords}   & $81,885$ & $4,748$ \\
$L_2$ & \makecell{Documents containing $\geq$20\% \\ of top 1\% co-occurring keywords} & $73,828$ & $1,785$ \\
$L_3$ & \makecell{Documents containing $\geq$10\% \\ of top 1\% co-occurring keywords} & $274,200$ & $112,030$ \\ \midrule

\rowcolor{gray!10} \multicolumn{2}{c}{\textbf{Total Documents}} & $433,016$ & $121,849$ \\ \bottomrule
         
\end{tabular}
}
\end{table}

\subsection{Feature Extraction}

DEEP transforms raw social movement discourse into structured temporal features across three complementary dimensions. Our feature extraction pipeline processes content from news sources and Reddit to generate time-indexed feature vectors that capture the evolving nature of social movement discourse. The complete details and mathematical formulation are provided in the Appendix.

\noindent\textbf{Volume Features ($\mathbf{V}_t$).} We quantify discourse intensity through platform-specific volume measurements that account for both content quantity and engagement patterns. Each content item is weighted by relevance scores derived from our multi-layer extraction methodology and engagement metrics. Volume features include raw counts, temporal derivatives (velocity and acceleration), and cross-platform distribution indices to capture how discourse spreads across news sources and Reddit channels.

\noindent\textbf{Emotion Features ($\mathbf{E}_t$).} We characterize the affective landscape using a 28-dimensional emotion space based on the GoEmotions dataset \cite{demszky-etal-2020-goemotions}. Content-level emotion intensities are computed using a fine-tuned RoBERTa model\footnote{\url{https://huggingface.co/SamLowe/roberta-base-go_emotions}}, then aggregated into weighted bin distributions across five intensity levels (Absent to Very High). This captures both the dominant emotional themes and their distributional patterns over time.

\noindent\textbf{Thematic Features ($\mathbf{T}_t$).} We capture the topical landscape through distance-based binning that leverages extracted keyphrases to quantify content alignment with predefined themes. Using a taxonomy of journalist-defined thematic categories (e.g., Gender Equality, Human Rights, Violence), we compute semantic distances between content keyphrases and topic centroids using DistilBERT embeddings \cite{DBLP:journals/corr/abs-1910-01108}, then discretize these distances into six bins representing alignment levels.

\noindent\textbf{Key Event Features ($\mathbf{f}_t^{\text{key}}$).} To model external catalysts, we incorporate journalist-identified events that may influence discourse patterns. Events are categorized across thematic dimensions and encoded with impact assessments relative to social movement objectives, generating binary indicator features for category-impact combinations.

\subsection{Transformer-based Forecasting}

DEEP leverages a time series transformer architecture \cite{DBLP:conf/ijcai/WenZZCMY023} to model the complex temporal dependencies inherent in social movement discourse evolution. Transformers are particularly well-suited for this task as they can capture long-range dependencies and handle multivariate time series with heterogeneous feature types. DEEP's forecasting component follows an encoder-decoder framework where the encoder processes historical discourse trajectories over $L_c$ context steps, and the decoder generates probabilistic predictions for $\Delta$ future time steps. 

The input representation at time $t$ integrates all discourse state components:
\begin{equation}
\mathbf{X}_t = [\mathbf{V}_t; \mathbf{E}_t; \mathbf{T}_t; \mathbf{f}_t^{\text{time}}; \mathbf{f}_t^{\text{key}}]
\end{equation}
where $\mathbf{V}_t \in \mathbb{R}^h$ encodes volume features, $\mathbf{E}_t \in \mathbb{R}^d$ captures emotional intensity distributions, $\mathbf{T}_t \in \mathbb{R}^m$ represents thematic distributions, $\mathbf{f}_t^{\text{time}} \in \mathbb{R}^2$ includes temporal encodings (day-of-week, month), and $\mathbf{f}_t^{\text{key}} \in \mathbb{R}^{Q \times (|\mathcal{I}| + 1)}$ encodes key event information from $\mathcal{K}_t$.

To maintain computational efficiency while preserving predictive information, we apply mutual information-based feature selection to reduce dimensionality before feeding inputs to the transformer layers.

Social movement discourse exhibits both immediate reactions and delayed cascading effects. To capture this multi-scale behavior, we incorporate explicit lag sequences $\mathcal{L} = \{1, 2, 3\}$ that provide the model direct access to recent historical states.

\emph{Rather than point predictions, DEEP generates full probability distributions over future discourse states.} We use a Student-t output distribution parameterized by location $\boldsymbol{\mu}_{t+\Delta}$, scale $\boldsymbol{\sigma}_{t+\Delta}$, and degrees of freedom $\boldsymbol{\nu}_{t+\Delta}$. This supports journalists' needs by providing both point forecasts and confidence intervals, enabling more informed reporting decisions when covering evolving social movements. 

\section{Experiments}

\subsection{Evaluation Metrics}
We evaluate DEEP's ability to produce accurate forecasts for journalists by predicting directional changes in target variables (e.g., volume of articles $\mathbf{V}_{t+\delta}$, or intensity of specific emotions within $\mathbf{E}_{t+\delta}$) over the $\delta$-day prediction horizon. For each target variable $Y_t$, we classify predictions into three categories using a statistical significance framework:
\begin{align}
\text{Increase: } &  Y_{t+\delta} > \mu_Y + 2\sigma_Y \\
\text{Decrease: } & Y_{t+\delta} < \mu_Y - 2\sigma_Y \\
\text{Stable: } & \mu_Y - 2\sigma_Y \leq Y_{t+\delta} \leq \mu_Y + 2\sigma_Y,
\end{align}
where $\mu_Y$ and $\sigma_Y$ represent the rolling mean and standard deviation computed over a 28-day historical window. The $2\sigma$ threshold corresponds to approximately 95\% confidence intervals, ensuring that predicted “significant" changes represent genuinely newsworthy developments rather than normal discourse fluctuation. 

Performance is measured using standard classification metrics across the three-class prediction task: accuracy, precision, recall and F1 scores, computed both per-class and macro-averaged. We report performance at different values of $\delta$, helping us understand the reliability of DEEP when considering longer horizons. 

\subsection{Implementation Details}

The TimeSeriesTransformer was configured with parameters chosen to balance journalists' needs and computational efficiency: (1) A 4-week historical window captures monthly discourse patterns while providing sufficient context for detecting emerging trends. (2) A 7-day prediction horizon ($\Delta = 7$ days) aligns with the weekly news cycle, enabling journalists to plan coverage, allocate resources, and prepare investigative pieces within standard editorial workflows. (3) A dropout rate of 0.3 mitigates overfitting, while a batch size of 2 accommodates the sequential nature of time-series data. (4) A learning rate of $3\times10^{-4}$ with a weight decay of $1\times10^{-5}$ supports stable convergence during training according to our empirical results.

Key events $\mathcal{K}_t$ were extracted from a comprehensive public source\footnote{\url{https://www.onthisday.com}} covering the study period from September 2024 to February 2025. Each event $k_i = (\epsilon_i, t_i, \alpha_i)$ was manually annotated by our team. \emph{The resulting annotated events table will be provided as part of the dataset when this paper is published.}
All experiments are conducted using three NVIDIA RTX A6000 GPUs with 48GB VRAM each. 

\subsection{DEEP System Performance}

\begin{table*}[t]
\centering
\renewcommand{\arraystretch}{0.9} 
\scalebox{0.78}{
\begin{tabular}{p{0.75cm} r|cc|cc|cc|cc|cc}
\toprule
& \textbf{Target} & 
\multicolumn{2}{c|}{\textbf{Precision@1}} & 
\multicolumn{2}{c|}{\textbf{Recall@1}} &
\multicolumn{2}{c|}{\textbf{F1@1}} & 
\multicolumn{2}{c}{\textbf{Accuracy@1}} &
\multicolumn{2}{c}{\textbf{AUC@1}} \\
& & Reddit & News & Reddit & News & Reddit & News & Reddit & News & Reddit & News \\
\midrule
\multirow{27}{*}{\rotatebox{90}{Emotion}} \\
& Admiration & 0.881 & 0.982 & 0.819 & 0.952 & 0.835 & 0.966 & 0.854 & 0.975 & 0.916 & 0.983 \\
& Amusement & 0.786 & 0.968 & 0.749 & 0.946 & 0.755 & 0.956 & 0.765 & 0.964 & 0.885 & 0.989 \\
& Anger & 0.801 & 0.954 & 0.759 & 0.929 & 0.768 & 0.940 & 0.776 & 0.954 & 0.887 & 0.989 \\
& Annoyance & 0.843 & 0.954 & 0.818 & 0.947 & 0.823 & 0.951 & 0.826 & 0.964 & 0.917 & 0.996 \\
& Approval & 0.701 & 0.778 & 0.664 & 0.776 & 0.670 & 0.774 & 0.680 & 0.783 & 0.834 & 0.917 \\
& Caring & 0.884 & 0.958 & 0.846 & 0.923 & 0.860 & 0.939 & 0.872 & 0.957 & 0.927 & 0.988 \\
& Confusion & 0.858 & 0.968 & 0.815 & 0.934 & 0.830 & 0.950 & 0.854 & 0.975 & 0.951 & 0.988 \\
& Curiosity & 0.779 & 0.901 & 0.742 & 0.871 & 0.755 & 0.883 & 0.754 & 0.886 & 0.902 & 0.970 \\
& Desire & 0.830 & 0.981 & 0.765 & 0.929 & 0.786 & 0.951 & 0.819 & 0.975 & 0.900 & 0.992 \\
& Disappointment & 0.813 & 0.961 & 0.794 & 0.877 & 0.794 & 0.911 & 0.811 & 0.936 & 0.892 & 0.970 \\
& Disapproval & 0.876 & 0.974 & 0.817 & 0.929 & 0.833 & 0.948 & 0.872 & 0.964 & 0.944 & 0.992 \\
& Disgust & 0.802 & 0.976 & 0.725 & 0.907 & 0.736 & 0.936 & 0.783 & 0.961 & 0.869 & 0.984 \\
& Embarrassment & 0.849 & 0.978 & 0.834 & 0.964 & 0.839 & 0.970 & 0.851 & 0.972 & 0.936 & 1.000 \\
& Excitement & 0.810 & 0.974 & 0.787 & 0.958 & 0.794 & 0.965 & 0.794 & 0.972 & 0.901 & 0.998 \\
& Fear & 0.829 & 0.951 & 0.801 & 0.932 & 0.811 & 0.941 & 0.819 & 0.950 & 0.936 & 0.994 \\
& Gratitude & 0.799 & 0.979 & 0.772 & 0.952 & 0.781 & 0.964 & 0.794 & 0.968 & 0.881 & 0.992 \\
& Grief & 0.936 & 0.965 & 0.942 & 0.912 & 0.939 & 0.935 & 0.940 & 0.947 & 0.986 & 0.989 \\
& Joy & 0.807 & 0.984 & 0.768 & 0.944 & 0.779 & 0.961 & 0.797 & 0.975 & 0.894 & 0.988 \\
& Love & 0.836 & 0.995 & 0.774 & 0.972 & 0.793 & 0.983 & 0.811 & 0.993 & 0.881 & 0.982 \\
& Nervousness & 0.853 & 0.962 & 0.840 & 0.845 & 0.843 & 0.884 & 0.861 & 0.947 & 0.916 & 0.987 \\
& Optimism & 0.901 & 0.975 & 0.888 & 0.956 & 0.894 & 0.964 & 0.893 & 0.964 & 0.957 & 0.999 \\
& Pride & 0.857 & 0.980 & 0.738 & 0.890 & 0.763 & 0.925 & 0.858 & 0.972 & 0.864 & 0.986 \\
& Realization & 0.781 & 0.856 & 0.729 & 0.830 & 0.744 & 0.840 & 0.765 & 0.868 & 0.864 & 0.931 \\
& Relief & 0.883 & 0.981 & 0.781 & 0.937 & 0.804 & 0.957 & 0.865 & 0.972 & 0.890 & 1.000 \\
& Remorse & 0.854 & 0.972 & 0.764 & 0.954 & 0.786 & 0.962 & 0.829 & 0.964 & 0.904 & 0.988 \\
& Sadness & 0.793 & 0.984 & 0.758 & 0.979 & 0.769 & 0.981 & 0.794 & 0.975 & 0.892 & 0.996 \\
& Surprise & 0.772 & 0.962 & 0.732 & 0.964 & 0.741 & 0.963 & 0.765 & 0.968 & 0.854 & 0.992 \\
\midrule
\rowcolor{gray!10} &  Average & $0.829_{0.047}$ & $0.950_{0.057}$ & $0.786_{0.054}$ & $0.916_{0.058}$ & $0.797_{0.052}$ & $0.930_{0.057}$ & $0.818_{0.051}$ & $0.944_{0.057}$ & $0.904_{0.033}$ & $0.981_{0.027}$ \\
\midrule

\multirow{3}{*}{\rotatebox{90}{Volume}} \\
& Raw & 0.884 & 0.864 & 0.833 & 0.783 & 0.852 & 0.811 & 0.856 & 0.817 & 0.869 & 0.803 \\
& Velocity & 0.656 & 0.799 & 0.597 & 0.779 & 0.612 & 0.788 & 0.612 & 0.791 & 0.756 & 0.811 \\
\midrule
\rowcolor{gray!10} & Average & $0.770_{0.114}$ & $0.832_{0.032}$ & $0.715_{0.118}$ & $0.781_{0.002}$ & $0.732_{0.120}$ & $0.799_{0.012}$ & $0.734_{0.122}$ & $0.804_{0.013}$ & $0.813_{0.021}$ & $0.807_{0.015}$\\
\bottomrule
\end{tabular}
}
\caption{Macro-averaged performance metrics (Precision, Recall, F1) and overall Accuracy and AUC across Reddit and News sources. The final row shows the mean and standard deviation across all emotions.}
\label{tab:emotion-metrics}
\end{table*}

\textbf{Next-day Prediction Performance.}
Table~\ref{tab:emotion-metrics} reports macro-average metrics at $\Delta = 1$ day for the three-way \emph{increase / stable / decrease} DEEP forecasting task on Reddit and news sources.

\noindent{\emph{DEEP performs markedly better on news than on Reddit. } Across emotions, precision varies from $0.829$ on Reddit to $0.95$ on news, recall from $0.786$ to $0.916$, F1 from $0.797$ to $0.93$, accuracy from $0.818$ to $0.944$, and AUC from $0.904$ to $0.981$. This gap likely reflects the more formal and topically focused language of news articles compared to the noisier and more informal expressions found on social media.

\noindent{\emph{DEEP's Performance varies by emotion but is generally strong.} Highly predictable emotions include \emph{Grief} (F1: $0.939$ Reddit, $0.935$ news) and \emph{Optimism} ($0.894$, $0.964$). Others improve dramatically from Reddit to news: \emph{Joy} ($0.779$ to $0.961$), \emph{Relief} ($0.804$ to $0.957$), and \emph{Amusement} ($0.755$ to $0.956$). Some emotions \emph{Disapproval} are almost perfectly predicted on news (precision = $0.974$, F1 = $0.948$).


\noindent{\emph{For some emotions, especially on Reddit, DEEP achieves high precision but lower recall.} For \emph{Nervousness}, DEEP achieves $0.920$ precision but $0.738$ recall on Reddit, indicating reliable but conservative predictions. This high-precision benefits journalists by minimizing false positives that could lead to investigating trends that never develop.\newline 


\noindent \textbf{One-Week Prediction Performance.}
We now  examine DEEP’s performance changes when $\Delta$ increases from 1 to 7 days. Figures~\ref{fig:reddits|increase} and~\ref{fig:reddits|decrease} report precision trends for the \emph{Increase} and \emph{Decrease} classes on Reddit, while Figures~\ref{fig:news|increase} and~\ref{fig:news|decrease} show the corresponding results for news. 

\noindent\emph{Reddit benefits more from longer horizons than news.} Reddit precision improves with increasing $\Delta$ for both \emph{Increase} and \emph{Decrease} classes, suggesting emotional shifts are more predictable over medium-term intervals.

\noindent\emph{News shows variable, emotion-specific horizon effects.} For \emph{Increase} predictions, some emotions have high precision (\emph{Admiration}) while others improve (\emph{Approval}, \emph{Sadness}). For \emph{Decrease} predictions, results are mixed: \emph{Approval} benefits from longer horizons while \emph{Admiration} declines, reflecting rapid news cycle shifts in emotional framing.



Overall, our findings indicate that DEEP achieves high precision on News even at short horizons, enabling journalists to act promptly on emerging narratives with confidence. Short-term forecasts are less reliable on Reddit, but predictability improves over longer horizons, making multi-day predictions valuable early indicators of slower-moving, community-driven trends. By leveraging both, journalists can combine immediate, trustworthy signals from news with horizon-informed insights from social media, allowing them to track and predict emotion shifts about SMs such as \#MeToo.

\begin{figure}[t]
     \centering
     \includegraphics[width=.65\linewidth]{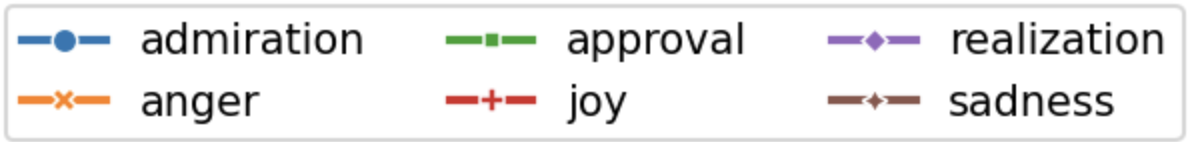}
     
     \subfloat[][Reddit - Increase]{\includegraphics[width=.24\textwidth]{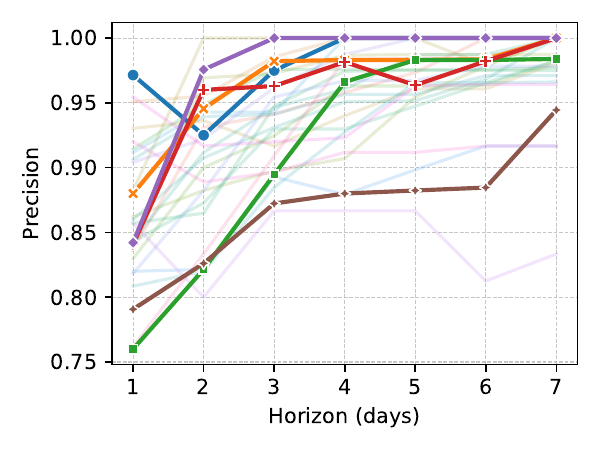}\label{fig:reddits|increase}}
     \subfloat[][Reddit - Decrease]{\includegraphics[width=.24\textwidth]{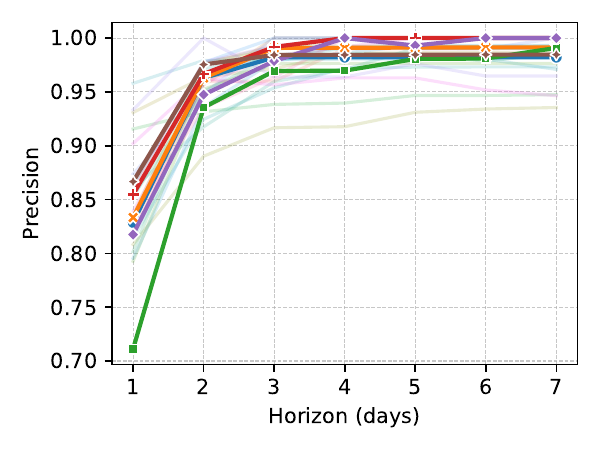}\label{fig:reddits|decrease}}
     
     \subfloat[][News - Increase]{\includegraphics[width=.24\textwidth]{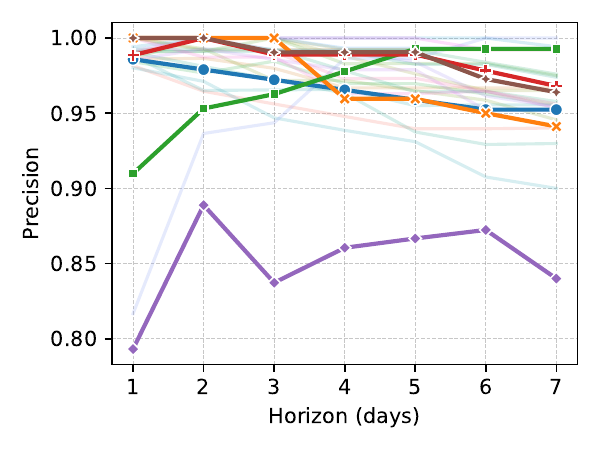}\label{fig:news|increase}}
     \subfloat[][News - Decrease]{\includegraphics[width=.24\textwidth]{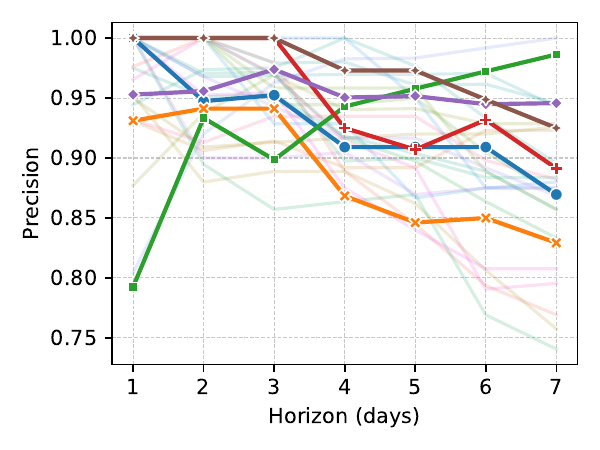}\label{fig:news|decrease}}     
    \caption{\textbf{Precision trends for forecasting emotional changes at varying horizons ($\Delta \in [1,7]$ days). }
    (a) Precision for Reddit, \emph{increase} class; 
    (b) Precision for Reddit, \emph{decrease} class; 
    (c) Precision for News, \emph{increase} class; 
    (d) Precision for News, \emph{decrease} class. }

     \label{fig:k_exps}
\end{figure}

\subsection{Real-World Case Study}

\begin{figure}[t]
    \centering
    \includegraphics[width=\linewidth]{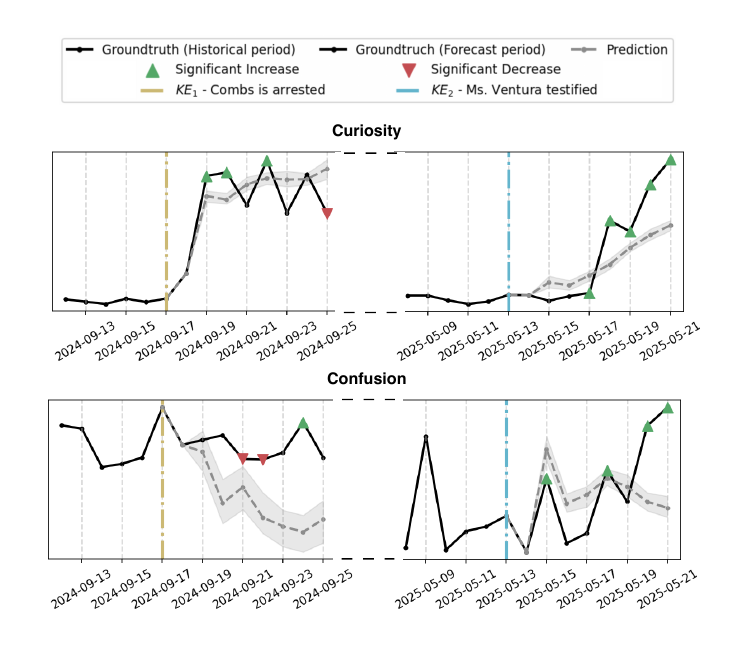}
    \caption{\textbf{Case study on Sean “Diddy” Combs}. The top panel shows the temporal trends for Curiosity and the bottom panel for Confusion. In both panels, the solid black line represents the ground truth, while the dashed gray line corresponds to DEEP’s forecast. Green and red markers indicate significant increases and decreases in the forecast period, respectively. Vertical lines denote the two key events: Combs' arrest on September 17, 2024 (yellow) and Ventura's courtroom testimony on May 13, 2025 (cyan).}
    \label{fig:casestudy}
\end{figure}

The allegations and legal proceedings involving Sean “Diddy” Combs represent a high-profile intersection between celebrity culture and \#MeToo\footnote{\url{https://www.bbc.com/news/articles/c869qd5j09xo}}. The timeline spans from Cassie Ventura’s initial lawsuit in November 2023, alleging years of abuse and coerced participation in drug-fueled “freak-offs”, through a cascade of lawsuits, the emergence of video evidence, and Combs' arrest on September 17, 2024. In the subsequent months, the trial brought forward multiple testimonies, including Ventura’s detailed account on May 13, 2025, which described coercion, violence, and psychological abuse. While the case ended with Combs being cleared of the most serious charges of racketeering and sex trafficking, he was convicted on two counts of transportation to engage in prostitution.

For this brief illustrative case study, we selected two pivotal key events: ($KE_1$) September 17, 2024 (Combs’ arrest) and ($KE_2$) May 13, 2025 (Ventura’s testimony). These events were chosen due to their potential to significantly alter both the volume and emotional tone of public conversation, particularly on social media platforms.

Figure \ref{fig:casestudy} shows the real and DEEP forecast trajectories of two emotions (curiosity and confusion) in Reddit discussions about \#MeToo SM (other emotions are in the Appendix). DEEP used historical data up to each key event (i.e., $KE_1$ and $KE_2$) to predict the subsequent week’s emotional trends. Ground truth values for the post-event period are plotted alongside predictions for comparison. Significant directional changes in emotion levels during the forecast windows are highlighted to emphasize shifts in discourse following each event.

The results demonstrate DEEP's ability to predict short-term emotional shifts in response to high-impact developments. Both key events were followed by a notable increase in \emph{Curiosity} within the discourse. DEEP's forecasts captured these upward trends, even though the absolute predicted values differed from the ground truth. From a journalist’s perspective, correctly predicting the \emph{direction} of change is more important than matching exact magnitudes, as it signals whether public interest is likely to grow or wane in days following a key event. Hence, DEEP’s provides timely, accurate trend prediction for newsroom decision-making.

For \emph{confusion}, a more nuanced pattern emerged. Following the arrest on September 17, 2024, confusion levels decreased, potentially reflecting a perceived clarification of the case’s trajectory with the formal filing of charges. In contrast, Ventura’s testimony on May 13, 2025 was followed by a marked increase in confusion, likely tied to the emotionally complex and contradictory narratives presented during trial. In both cases, DEEP accurately predicted the direction of change, underscoring its ability to model how the interplay of new evidence, legal framing, and public perception shapes online emotional dynamics.

\section{Related Work}

Prior research \cite{ng2023experimental,10.1145/3708513,He_Shen_Mukherjee_Vucetic_Dragut_2021} on forecasting social-media discourse has largely focused on predicting activity volume, often leveraging exogenous real-world event signals such as news, GDELT\footnote{\url{https://www.gdeltproject.org}}, or ACLED\footnote{\url{https://acleddata.com}}, but rarely extending to sentiment or discrete emotion forecasting, particularly in a multi-platform setting involving Reddit and news sources. Examples include \emph{ChatterNet}~\cite{10.1145/3394486.3403251}, which predicts subreddit “chatter” volume by combining contemporaneous news-derived features with  Reddit activity, and the \emph{TAP} family of models~\cite{DBLP:journals/corr/abs-2109-11024,ng2022social}, which use LSTMs to forecast topic-specific daily volumes over short horizons. Similar approaches on Twitter~\cite{Yates_Joselow_Goharian_2021} model news-driven spikes and demonstrate the benefits of exogenous signals for event-aligned bursts in volume. Work on affect dynamics \cite{doi:10.1177/1754073915590619,sener2023unveiling,10.1145/2346676.2346682} constructs discrete emotion time series and uses transformer-based topic modeling to explain shifts (e.g., during Black Lives Matter \cite{10.1145/3625007.3627477}), but stops short of forecasting. Sentiment is often used as an input feature~\cite{saleiro2016learning,iamnitchi2023modeling} but predicting future sentiment has been rare. Across this literature, forecasting is dominated by RNN/GRU/LSTM architectures, and transformer use is limited to topic modeling rather than temporal prediction. No prior work jointly forecasts volume and discrete emotions across Reddit and news in a coordinated, multi-output framework, a gap that our DEEP framework addresses.

\section{Conclusion}

This paper presents the Discourse Evolution Engine for Predictions about Social Movements (DEEP) system that enables journalists to forecast volume and emotional changes in social movement discourse across Reddit and news sources. Through experiments on \#MeToo data, DEEP achieves strong performance (e.g., F1 scores averaging 0.912 on news, 0.745 on Reddit for emotions classification) with complementary predictive patterns: high precision on news at short horizons and improved Reddit predictions over longer horizons. While designed for social movements, our keyword-based framework generalizes to other topic-based forecasting applications. Developed through engagement with over 20 journalists worldwide, DEEP addresses the need to identify emerging trends before mainstream attention, enabling journalists to anticipate discourse shifts and make more informed editorial decisions.

\section{Acknowledgements}

This research was supported by the Roberta Buffett Institute for Global Affairs at Northwestern University. Additional support was provided by Amazon Web Services (AWS) through the AWS Cloud Credits for Research program. 

\bibliography{aaai25}

\clearpage

\appendix

\section{DEEP Participating Organizations}

To ensure that SMART/DEEP serves as an effective tool for journalists seeking to understand the evolution and community impact of social movements, we engaged in a comprehensive stakeholder consultation process involving multiple phases of input and feedback.

\paragraph{Initial Stakeholder Interviews}
We conducted in-depth interviews with eight journalists representing four diverse news organizations: Grist (2 journalists), The 19th (1 journalist), The Associated Press (3 journalists), and The Examination (2 journalists). These interviews served multiple purposes: first, to establish working definitions of “social movement" from the journalistic perspective; second, to identify current methodologies journalists employ for tracking social movements; and third, to document existing pain points and workflow challenges in their reporting processes. Based on these insights, we collaboratively brainstormed potential features and functionalities for the SMART/DEEP platform.

\paragraph{Prototype Evaluation and Refinement}
Following the initial development phase, we returned to the same cohort of journalists to demonstrate the SMART/DEEP prototype. This iterative feedback process allowed us to collect targeted input on the tool's functionality, user interface, and overall efficacy, leading to substantial refinements in the platform's design and capabilities.

\paragraph{Extended Workshop Validation}
To broaden our stakeholder engagement and validate findings across a more diverse journalism ecosystem, we organized a workshop in March 2025\footnote{\url{https://buffett.northwestern.edu/news/2025/empowering-journalists-to-decode-social-movements-the-buffett-global-working-group-behind-new-ai-system.html}}. This workshop included representatives from nine additional organizations, encompassing both traditional and digital media outlets: The Wall Street Journal, American Press Institute, LocalMedia Association, The Washington Post, Borderless Magazine, Institute for Nonprofit News, ChicagoPublicMedia, JournalismAI, and Chicago Sun-Times Media. 

\section{Journalist-defined Search Keywords}

The keyword set defined by journalists for our daily search of news articles and Reddit posts is listed as follows: \textit{black feminism, child marriage, domestic violence, equal opportunities, equal representation in the workplace, female education, female empowerment, female genital mutilation, feminism, gender bias, gender discrimination, gender equality, gender equality in politics, gender equity, gender expression, gender gap, gender identity, gender inequality, gender justice, gender norms, gender parity, gender pay gap, gender-responsive budgeting, gender roles, gender-sensitive, gender-sensitive education, gender-sensitive healthcare, gender-sensitive policies, gender stereotypes, maternal health, maternal mortality, \#MeToo, patriarchy, reproductive health, reproductive rights, sexual harassment, sexual reproduction and health rights, violence against women, women in leadership, women in politics, women in power, women's autonomy, women's economic empowerment, women's empowerment, women's entrepreneurship, women's health, women's land rights, women's leadership, women's participation, women's representation, women's rights, women's safety, women's voices.}

\section{Feature Extraction}

\subsubsection{Volume Features}
The discourse volume $\mathbf{V}_t$ represents the volume of relevant posts/articles (henceforth referred to as ``content items'') at time $t$. We decompose $V_t$ into platform-specific components and temporal aggregation functions to capture the multifaceted nature of social movement discourse dissemination. Each component is explained in detail as follows.

\noindent\textit{Platform-Specific Volume Components. }
Let $\mathcal{P} = \{P_1, P_2, \ldots, P_n\}$ represent the set of monitored platforms, where each  $P_i$ is a social platform (e.g., X, Facebook, Reddit), traditional news outlet, or other information channel. The volume contribution from platform $P_i$ at time $t$ is denoted as:
\begin{equation}
V_{i,t} = \sum_{j=1}^{N_{i,t}} w_{i,j,t} \cdot \rho_{i,j,t},
\end{equation}
where $N_{i,t}$ is the number of relevant content items on platform $P_i$ at time $t$, $w_{i,j,t}$ represents the engagement weight for content item $j$ (e.g., number of likes, shares, comments, or readership), and $\rho_{i,j,t}\in[0,1]$ is the relevance score indicating how closely content item $j$ aligns with the social movement discourse.
The relevance score $\rho_{i,j,t}$ is computed based on the multi-layer data extraction methodology. Specifically, for a content item $j$ classified into layer $L_\ell$ where $\ell \in \{0, 1, 2, 3\}$, the relevance score is defined as $\rho_{i,j,t}=1-\frac{\ell}{L_{\text{max}}}$, where $L_{\text{max}} = 3$ represents the maximum layer index in our implementation. \newline

\noindent\textit{Temporal Smoothing and Normalization}
To address the inherent noise in raw volume measurements and ensure temporal consistency, we apply a temporal smoothing function:
\begin{equation}
\tilde{V}_{i,t} = \sum_{\tau=0}^{W-1} \lambda^\tau \cdot V_{i, t-\tau} \cdot \frac{1-\lambda}{1-\lambda^W},
\end{equation}
where $W$ is the smoothing window size and $\lambda \in (0,1)$ is the exponentially decaying weight parameter such that more recent observations receive higher weights. The window size is set to $W = 7$ days to capture weekly patterns while maintaining responsiveness to emerging events. The decay parameter $\lambda = 0.8$ is chosen to balance recent information (giving higher weight to more recent days) with historical context. \newline 

\noindent\textit{Volume Intensity Measures. }
Beyond absolute volume, we define several derived intensity measures:
\begin{align}
\text{Volume Velocity: } & \quad \dot{V}_{i,t} = V_{i,t} - V_{i,t-1} \\
\text{Volume Acceleration: } & \quad \ddot{V}_{i,t} = \dot{V}_{i,t} - \dot{V}_{i,t-1} \\
\text{Standardized Volume: } & \quad V_{i,t}^{\text{std}} = \frac{V_{i,t} - \mu_{V_i}}{\sigma_{V_i}}
\end{align}
where $\mu_{V_i}$ and $\sigma_{V_i}$ represent the rolling mean and standard deviation of volume over a baseline period, enabling detection of significant deviations from typical discourse patterns. \newline 

\noindent\textit{Platform Distribution Index. }
To capture the distribution of discourse across platforms, we define a distribution index:
\begin{equation}
\text{PDI}_t = -\sum_{i=1}^{n} \pi_{i,t} \log(\pi_{i,t})
\end{equation}
where $\pi_{i,t} = \frac{V_{i,t}}{\sum_{j=1}^{n} V_{j,t}}$ is the proportional contribution of platform $P_i$ to total volume at time $t$. Higher values of $\text{PDI}_t$ indicate more distributed discourse across platforms, while lower values suggest focus on specific channels.

\subsubsection{Emotion Features}
The emotion component $\mathbf{E}_t$ captures the affective landscape of social movement discourse. We use a $d=28$-dimensional emotion space $\mathbf{E}_t \in \mathbb{R}^d$, with each dimension $\xi$ representing the intensity of emotion $\xi$ at time $t$. Our implementation uses all the $d = 28$ emotions from the GoEmotions dataset \cite{demszky-etal-2020-goemotions} (e.g., joy, anger, admiration, disappointment). \newline

\noindent\textit{Emotion Bins. } For each emotion dimension $\xi$, we discretize the continuous intensity space $[0, 1]$ into $B$ ordered bins. Using intensity-based quantile partitioning, we define:
\begin{equation}
\mathcal{B} = {[0, 0.2), [0.2, 0.4), [0.4, 0.6), [0.6, 0.8), [0.8, 1.0]}
\end{equation}
We use $B = 5$ bins per emotion dimension corresponding to: { Absent, Low, Moderate, High, Very High }. \newline 

\noindent\textit {Content-Level Emotion Assignment. }
For each content item $c_{i,j,t}$ from platform $P_i$ at time $t$, we compute raw emotion intensity scores $\mathbf{e}_{i,j,t} = [e_{i,j,t}^{(1)}, e_{i,j,t}^{(2)}, \ldots, e_{i,j,t}^{(d)}]^T$ using a RoBERTa architecture fine-tuned on the GoEmotions dataset \cite{demszky-etal-2020-goemotions}\footnote{\url{https://huggingface.co/SamLowe/roberta-base-go_emotions}} for emotion detection. The bin assignment function for emotion dimension $\xi$ is:
\begin{equation}
\beta_{\xi}(e_{i,j,t}^{(\xi)}) = b \text{ if } e_{i,j,t}^{(\xi)} \in \mathcal{B}_b
\end{equation}
\newline
\noindent\textit{Weighted Bin Distributions}
The bin-based emotion distribution for dimension $\xi$ at time $t$ is:
\begin{equation}
\mathbf{G}_{\xi,t} = [G_{\xi,1,t}, G_{\xi,2,t}, \ldots, G_{\xi,B,t}]^T
\end{equation}
where each bin probability is computed as:
\begin{equation}
G_{\xi,b,t} = \frac{\sum_{i=1}^{n} \sum_{j: \beta_{\xi}(e_{i,j,t}^{(\xi)}) = b} w_{i,j,t} \cdot \rho_{i,j,t}}{\sum_{i=1}^{n} \sum_{j=1}^{N_{i,t}} w_{i,j,t} \cdot \rho_{i,j,t}}
\end{equation}
This ensures $\sum_{b=1}^{B} G_{\xi,b,t} = 1$ for each emotion $\xi$. \newline 

\noindent\textit{Emotion Intensity Aggregation. }
From the bin distributions, we derive emotion-specific intensity measures:
\begin{align}
\text{Mean Intensity: }  \quad \bar{E}_{\xi,t} = \sum_{m=1}^{B} G_{\xi,m,t} \cdot \frac{b_{\xi,m-1} + b_{\xi,m}}{2} \\
\text{Emotion Variance: } \quad \sigma_{E_{\xi}}^2(t) = \sum_{m=1}^{B} G_{\xi,m,t} \cdot \\ \notag  \qquad\qquad \cdot\left(\frac{b_{\xi,m-1} + b_{\xi,m}}{2} - \bar{E}_{\xi,t}\right)^2 \\
\text{Peak Intensity: }  \quad \text{Peak}_{\xi,t} = \arg\max_m G_{\xi,m,t}
\end{align}
\noindent\textit{Emotion Concentration and Dispersion. }
We measure how concentrated emotions are across bins:
\begin{align}
\text{Concentration: } & \quad C_{\xi,t} = \sum_{m=1}^{B} G_{\xi,m,t}^2 \\
\text{Entropy: } & \quad H_{\xi,t} = -\sum_{m=1}^{B} G_{\xi,m,t} \log G_{\xi,m,t}
\end{align}
Higher concentration indicates focused emotional responses, while higher entropy suggest more unequal distribution of emotional intensity. \newline 

\noindent\textit{Cross-Emotion Correlation Analysis}
To capture emotion co-occurrence patterns, we compute pairwise emotion correlations using the Pearson correlation coefficient on mean intensities:
\begin{equation}
\text{Corr}(\xi,\xi',t) = \frac{\text{Cov}(\bar{E}_{\xi,t-W:t}, \bar{E}_{\xi',t-W:t})}{\sqrt{\text{Var}(\bar{E}_{\xi,t-W:t}) \cdot \text{Var}(\bar{E}_{\xi',t-W:t})}}
\end{equation}
where the covariance and variance are computed over a sliding window of size $W=7$.

\subsubsection{Thematic Features}
Thematic features $\mathbf{T}_t$ capture the topical landscape of social movement discourse via distance-based binning that leverages extracted keyphrases to quantify content alignment with predefined themes. \newline 

\noindent\textit{Topic Taxonomy and Keyphrases. }
We define a taxonomy of $L$ thematic categories relevant to the target social movement:
\begin{equation}
\mathcal{T} = \{T_1, T_2, \ldots, T_{L}\}
\end{equation}
Our journalist co-authors decided these include $L = 9$ topics\footnote{These topics can be adapted for different social movements by journalists familiar with the movement's context.}: Gender Equality, Human Rights, Violence, Health \& Reproductive Rights, Political Change, Natural Disaster, Climate \& Environment, Migration \& Displacement, and Technology \& AI. Each topic has an associated set of keyphrases (e.g.,  for Gender Equality we included the following keyphrases: “legalize same-sex marriage", “transgender", “women's rights", “female", “mayor", “LGBT", “gay", “pride parade").

For each content item $c_{i,j,t}$ from platform $P_i$ at time $t$, we have a set of keyphrases $\mathcal{K}_{i,j,t} = \{k_1, k_2, \ldots, k_{|\mathcal{K}{i,j,t}|}\}$. Each keyphrase $k_m$ is represented as a dense vector embedding $\mathbf{v}_m \in \mathbb{R}^D$ using DistilBERT ~\cite{DBLP:journals/corr/abs-1910-01108}. Each topic $T_\ell$ is represented by a centroid vector $\mathbf{c}_\ell \in \mathbb{R}^D$ computed as the average of keyphrases for that topic:
\begin{equation}
\mathbf{c}_\ell = \frac{1}{|\mathcal{R}_\ell|}\sum_{k \in \mathcal{R}_\ell}  \mathbf{v}_k,
\end{equation}
where $\mathcal{R}_\ell$ is the set of keyphrases for topic $T_\ell$. \newline

\noindent\textit{Distance Computation and Binning. }
For each keyphrase $k_m$ in content $c_{i,j,t}$, we compute the semantic distance to each topic centroid:
\begin{equation}
d(k_m, T_\ell) = 1 - \frac{\mathbf{v}_m \cdot \mathbf{c}_\ell}{|\mathbf{v}_m| |\mathbf{c}_\ell|}
\end{equation}
This cosine-based distance measure yields $d(k_m, T_\ell) \in [0, 2]$, where smaller values indicate greater thematic alignment. 

We discretize the distance space into $N = 6$ bins for each topic:
\begin{equation}
\mathcal{D} = {[0, 0.1), [0.1, 0.2), [0.2, 0.4), [0.4, 0.6), [0.6, 0.8), [0.8, 1.0]},
\end{equation}
corresponding to: { Core, Close, Related, Peripheral, Distant, Unrelated}. \newline 

\noindent\textit{Content-Level Thematic Profiles. }
For each content item $c_{i,j,t}$, we construct a thematic profile matrix $\mathbf{M}{i,j,t} \in \mathbb{R}^{L \times N}$ where element $M_{i,j,t}^{(\ell,n)}$ represents the proportion of keyphrases in distance bin $n$ for topic $T_\ell$:
\begin{equation}
M_{i,j,t}^{(\ell,n)} = \frac{|{k_m \in \mathcal{K}_{i,j,t} : d(k_m, T_\ell) \in \mathcal{D}_n}|}{|\mathcal{K}_{i,j,t}|}
\end{equation}
This ensures $\sum_{n=1}^{N} M_{i,j,t}^{(\ell,n)} = 1$ for each topic $T_\ell$. \newline 

\noindent\textit{Weighted Aggregate Thematic Distributions. }
The discourse-level thematic distribution for topic $T_\ell$ and distance bin $n$ at time $t$ is:
\begin{equation}
\Theta_{\ell,n,t} = \frac{\sum_{i=1}^{n} \sum_{j=1}^{N_{i,t}} w_{i,j,t} \cdot \rho_{i,j,t} \cdot M_{i,j,t}^{(\ell,n)}}{\sum_{i=1}^{n} \sum_{j=1}^{N_{i,t}} w_{i,j,t} \cdot \rho_{i,j,t}}
\end{equation}
We also measure topic competition through mutual information:
\begin{equation}
\text{MI}(\ell,\ell',t) = \sum_{h=1}^{N} \sum_{z=1}^{N} \Phi_{\ell,\ell'}(h,z,t) \log \frac{\Phi_{\ell,\ell'}(h,z,t)}{\Theta_{\ell,h,t} \cdot \Theta_{\ell',z,t}}
\end{equation}
where $\Phi_{\ell,\ell'}(h,z,t)$ is the joint probability of a content item having keyphrases in bin $h$ for topic $T_\ell$ and bin $z$ for topic $T_{\ell'}$.

\subsubsection{Key Event Features}
DEEP also supports journalists who believe that a key event is linked to shifts in discourse about an SM. The key event component $\mathcal{K}_t$ captures this. The key events we consider are related to the $L$ topics used to extract \emph{Thematic Features}. Relevant event information can be obtained from publicly available sources\footnote{\url{https://www.onthisday.com}}. Additional implementation details are provided in the Experiments section.


For each key event occurring at time $t$, we define an \emph{impact vector} that captures both categorical membership and directional influence on the social movement. The impact assessment is formalized through a mapping function $\iota: \mathcal{E} \times \mathcal{T}_{t} \rightarrow \mathcal{I}$, where $\mathcal{T}_{t}$ represents the set of events occurring at time $t$ and $\mathcal{I} = \{-1, 0, 1, 2\}$ denotes the impact scale. In the case of our \#MeToo example, we used:
$$\mathcal{I} = \begin{cases}
2 & \text{Opposes objectives of $sm$} \\
1 & \text{Supports objectives of $sm$} \\
0 & \text{Neutral impact} \\
-1 & \text{Not related to $sm$}
\end{cases}$$

For each event category $E_q$ and impact level $\iota \in \mathcal{I}$, we construct indicator variables that capture the presence and nature of events affecting discourse at time $t$. The binary encoding generates feature vectors $\boldsymbol{\Phi}_t \in \{0,1\}^{Q \times (|\mathcal{I}| + 1)}$ where each element corresponds to a category-impact combination. Specifically, for category $E_q$ and impact level $\iota$, the binary indicator is defined as:
$$\phi_{q,\iota,t} = \begin{cases}
1 & \text{if } \exists \text{ event in category } E_q \text{ with impact } \iota \text{ at time } t \\
0 & \text{otherwise}
\end{cases}$$

Additionally, our “Not Available" indicator $\phi_{q,\text{NA},t}$ for each category explicitly models the absence of events:
$$\phi_{q,\text{NA},t} = \begin{cases}
1 & \text{if no events in category } E_q \text{ occur at time } t \\
0 & \text{otherwise}
\end{cases}$$

\section{Training Dynamics}

Figure~\ref{fig:training_dynamics} reports the Negative Log-Likelihood (NLL) loss and Mean Squared Error (MSE) at horizon $\Delta = 1$ over training epochs. Both metrics show a clear decreasing trend, indicating that the time-series transformer progressively improves its ability to perform the forecasting task.

\begin{figure}
    \centering
    \includegraphics[width=0.9\linewidth]{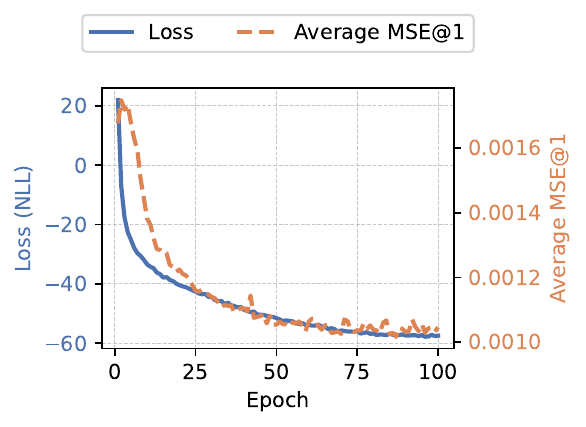}
    \caption{Training Dynamics: Negative Log-Likelihood (NLL) loss and mean squared error (MSE) determined at horizon $\Delta=1$ across different training epochs. }
    \label{fig:training_dynamics}
\end{figure}

\section{Case Study}

Table \ref{tab:all-emotions-forecast} shows the observed trend (i.e., ground truth) of each emotion following the two key events in our case study: (i) $KE_1$ Combs' arrest and (ii) $KE_2$ Ventura's testomony. In addition, we report whether DEEP predicted the correct trend for both Reddit and News.

\begin{table*}[t]
\centering
\caption{Summary of DEEP forecasting performance for all 28 emotions on Reddit and News for the two key events ($KE_1$: 17 September 2024, $KE_2$: 13 May 2025). Columns indicate whether the ground truth (GT) trend was “increase" (\uparrowtrend), “decrease" (\downarrowtrend), or “stable" (\stabletrend) and whether DEEP correctly predicted the trend (\correcttrend, \wrongtrend). Last row is the percentage of times DEEP predicted the correct GT.}
\label{tab:all-emotions-forecast}
\begin{tabular}{r|cc cc|cc cc}
\toprule
\multirow{3}{*}{\textbf{Emotion}} 
    & \multicolumn{4}{c}{$\mathbf{KE_1}$ (September 17, 2024)} 
    & \multicolumn{4}{c}{$\mathbf{KE_2}$ (May 13, 2025)} \\
\cmidrule(lr){2-5} \cmidrule(lr){6-9}
    & \multicolumn{2}{c}{\textbf{Reddit}} 
    & \multicolumn{2}{c}{\textbf{News}} 
    & \multicolumn{2}{c}{\textbf{Reddit}} 
    & \multicolumn{2}{c}{\textbf{News}} \\
\cmidrule(lr){2-3} \cmidrule(lr){4-5} \cmidrule(lr){6-7} \cmidrule(lr){8-9}
    & GT &  DEEP & GT & DEEP & GT & DEEP & GT & DEEP \\
\midrule
admiration      & \uparrowtrend   & \correcttrend  & \stabletrend    & \correcttrend    &  \uparrowtrend   & \wrongtrend    &  \stabletrend    & \correcttrend  \\
amusement       & \stabletrend    & \correcttrend  & \stabletrend    & \correcttrend    &  \uparrowtrend   & \wrongtrend    &  \stabletrend    & \wrongtrend  \\
anger           & \uparrowtrend   & \wrongtrend    & \stabletrend    & \wrongtrend      &  \uparrowtrend   & \correcttrend  &  \stabletrend    & \wrongtrend  \\
annoyance       & \uparrowtrend   & \correcttrend  & \stabletrend    & \correcttrend    &  \downarrowtrend & \correcttrend  &  \stabletrend    & \correcttrend  \\
approval        & \stabletrend    & \correcttrend  & \uparrowtrend   & \correcttrend    &  \uparrowtrend   & \wrongtrend    &  \uparrowtrend   & \correcttrend  \\
caring          & \stabletrend    & \correcttrend  & \stabletrend    & \correcttrend    &  \downarrowtrend & \correcttrend  &  \stabletrend    & \correcttrend  \\
confusion       & \downarrowtrend & \correcttrend  & \stabletrend    & \wrongtrend      &  \uparrowtrend   & \correcttrend  &  \downarrowtrend & \correcttrend  \\
curiosity       & \uparrowtrend   & \correcttrend  & \downarrowtrend & \correcttrend    &  \uparrowtrend   & \correcttrend  &  \stabletrend    & \wrongtrend  \\
desire          & \downarrowtrend & \correcttrend  & \uparrowtrend   & \correcttrend    &  \uparrowtrend   & \correcttrend  &  \stabletrend    & \wrongtrend  \\
disappointment  & \uparrowtrend   & \correcttrend  & \uparrowtrend   & \correcttrend    &  \uparrowtrend   & \correcttrend  &  \uparrowtrend   & \wrongtrend  \\
disapproval     & \uparrowtrend   & \correcttrend  & \stabletrend    & \correcttrend    &  \downarrowtrend & \correcttrend  &  \stabletrend    & \wrongtrend  \\
disgust         & \downarrowtrend & \correcttrend  & \stabletrend    & \wrongtrend      &  \uparrowtrend   & \wrongtrend    &  \stabletrend    & \correcttrend  \\
embarrassment   & \stabletrend    & \correcttrend  & \stabletrend    & \wrongtrend      &  \uparrowtrend   & \wrongtrend    &  \stabletrend    & \correcttrend  \\
excitement      & \uparrowtrend   & \correcttrend  & \stabletrend    & \wrongtrend      &  \uparrowtrend   & \correcttrend  &  \stabletrend    & \wrongtrend  \\
fear            & \uparrowtrend   & \wrongtrend    & \stabletrend    & \wrongtrend      &  \uparrowtrend   & \correcttrend  &  \downarrowtrend & \correcttrend  \\
gratitude       & \stabletrend    & \wrongtrend    & \stabletrend    & \wrongtrend      &  \downarrowtrend & \correcttrend  &  \stabletrend    & \correcttrend  \\
grief           & \stabletrend    & \correcttrend  & \stabletrend    & \correcttrend    &  \uparrowtrend   & \correcttrend  &  \stabletrend    & \correcttrend  \\
joy             & \uparrowtrend   & \correcttrend  & \downarrowtrend & \correcttrend    &  \stabletrend    & \correcttrend  &  \stabletrend    & \wrongtrend  \\
love            & \uparrowtrend   & \correcttrend  & \stabletrend    & \wrongtrend      &  \uparrowtrend   & \correcttrend  &  \stabletrend    & \wrongtrend  \\
nervousness     & \uparrowtrend   & \correcttrend  & \stabletrend    & \wrongtrend      &  \uparrowtrend   & \wrongtrend    &  \stabletrend    & \wrongtrend  \\
optimism        & \uparrowtrend   & \wrongtrend    & \stabletrend    & \correcttrend    &  \uparrowtrend   & \correcttrend  &  \stabletrend    & \wrongtrend  \\
pride           & \stabletrend    & \correcttrend  & \stabletrend    & \wrongtrend      &  \stabletrend    & \correcttrend  &  \uparrowtrend   & \wrongtrend  \\
realization     & \downarrowtrend & \correcttrend  & \uparrowtrend   & \correcttrend    &  \stabletrend    & \correcttrend  &  \stabletrend    & \wrongtrend  \\
relief          & \stabletrend    & \wrongtrend    & \stabletrend    & \wrongtrend      &  \stabletrend    & \wrongtrend    &  \stabletrend    & \wrongtrend  \\
remorse         & \stabletrend    & \wrongtrend    & \stabletrend    & \wrongtrend      &  \uparrowtrend   & \wrongtrend    &  \stabletrend    & \wrongtrend  \\
sadness         & \uparrowtrend   & \wrongtrend    & \stabletrend    & \correcttrend    &  \uparrowtrend   & \wrongtrend    &  \uparrowtrend   & \wrongtrend  \\
surprise        & \uparrowtrend   & \correcttrend  & \stabletrend    & \wrongtrend      &  \uparrowtrend   & \correcttrend  &  \uparrowtrend   & \wrongtrend  \\ \midrule

\rowcolor{gray!10}Average & \multicolumn{2}{c}{$74.1\%$} & \multicolumn{2}{c}{$51.8\%$} & \multicolumn{2}{c}{$62.9\%$} & \multicolumn{2}{c}{$37.1\%$} \\
\bottomrule
\end{tabular}
\end{table*}

\end{document}